\documentclass[prb,twocolumn,showpacs,amsmath,amssymb]{revtex4}

\usepackage{graphicx}

\usepackage{enumerate}
\usepackage{longtable}

\usepackage{dcolumn}
\usepackage{bm}
\usepackage{color}

\begin{document}
\title{Mott-Hubbard exciton in the optical conductivity of YTiO$_3$ and SmTiO$_3$}
\author{A.~G\"{o}ssling$^{1}$}
\author{R.~Schmitz$^2$}
\author{H.~Roth$^1$}
\author{M.W.~Haverkort$^1$}
\author{T.~Lorenz$^1$}
\author{J.A.~Mydosh$^1$}
\author{E.~M\"{u}ller-Hartmann$^2$}
\author{M.~Gr\"{u}ninger$^{1,3}$}
\affiliation{$^1$II. Physikalisches Institut,
Universit\"{a}t zu K\"{o}ln, Z\"{u}lpicher Strasse 77, D-50937 K\"{o}ln, Germany\\
$^2$Institut f\"{u}r Theoretische Physik, Universit\"{a}t zu K\"{o}ln,
Z\"{u}lpicher Strasse 77, D-50937 K\"{o}ln, Germany\\
$^3$2.\ Physikalisches Institut A, RWTH Aachen University,
Otto-Blumenthal-Strasse, D-52074 Aachen, Germany}
\date{May 16, 2008}

\begin{abstract}
In the Mott-Hubbard insulators YTiO$_3$ and SmTiO$_3$ we study optical excitations from the lower
to the upper Hubbard band, $|d^1d^1\rangle \! \rightarrow \! |d^0d^2\rangle$.
The multi-peak structure observed in the optical conductivity reflects the multiplet structure of
the upper Hubbard band in a multi-orbital system.
Absorption bands at 2.55 and 4.15\,eV in the ferromagnet YTiO$_3$ correspond to final states
with a triplet $d^2$ configuration, whereas a peak at 3.7\,eV in the antiferromagnet SmTiO$_3$
is attributed to a singlet $d^2$ final state.
A strongly temperature-dependent peak at 1.95\,eV in YTiO$_3$ and 1.8\,eV in SmTiO$_3$ is interpreted
in terms of a Hubbard exciton, i.e., a charge-neutral (quasi-)bound state of a hole in the lower Hubbard
band and a double occupancy in the upper one.
The binding to such a Hubbard exciton may arise both due to Coulomb attraction
between nearest-neighbor sites and due to a lowering of the kinetic energy
in a system with magnetic and/or orbital correlations.
Furthermore, we observe anomalies of the spectral weight in the vicinity of the magnetic
ordering transitions, both in YTiO$_3$ and SmTiO$_3$. In the $G$-type antiferromagnet SmTiO$_3$,
the {\it sign} of the change of the spectral weight at $T_N$ depends on the polarization.
This demonstrates that the temperature dependence of the spectral weight is not dominated by
the spin-spin correlations, but rather reflects small changes of the orbital occupation.
\end{abstract}

\maketitle

\section{Introduction}

\begin{figure}[tb]
\includegraphics[width=0.9\columnwidth,clip]{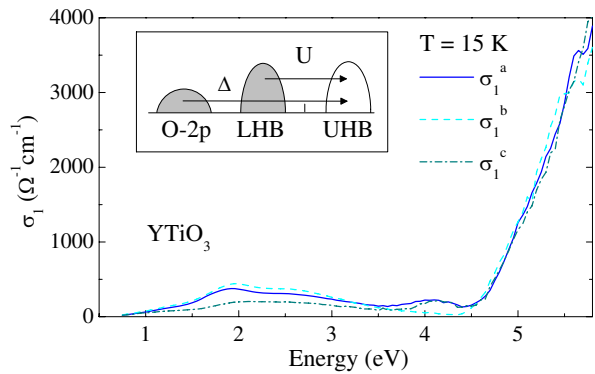}
\caption{(Color online) Optical conductivity of YTiO$_3$ at 15\,K.\@
   \emph{Inset:} sketch of the optical excitations from the lower Hubbard
   band (LHB) and the oxygen $2p$ band into the upper Hubbard band (UHB)
   in case of a single, half-filled orbital at the transition-metal site. }
\label{fig:Hubbard}
\end{figure}

In strongly correlated electron systems, the competition between kinetic energy
and Coulomb repulsion gives rise to a variety of intriguing phenomena.\cite{kotliar04a,imada98a}
The most simple approach is the single-band Hubbard model, with on-site Coulomb repulsion $U$
and a kinetic part proportional to the inter-site hopping amplitude $t$. For $U$ larger than the
band width, the band splits into a lower and an upper Hubbard band (LHB and UHB, see inset of
Fig.\ \ref{fig:Hubbard}). At half filling one finds a Mott-Hubbard insulator with one localized electron per site,
i.e., the Coulomb energy dominates. However, the low-energy physics is determined by the kinetic
energy: virtual hopping of the electrons to neighboring sites is effectively described by exchange
interactions with $J \propto t^2/U$. These govern the spin degrees of freedom and, in a multi-orbital model,
are also relevant for the orbital degrees of freedom.\cite{tokura00,khaliullinrev,khomskiirev}

The competition between Coulomb energy and kinetic energy also governs the formation of bound states,
e.g., excitons. In simple band insulators, binding of an electron in the conduction band and a hole in the
valence band reduces the Coulomb energy, while the kinetic energy increases.
In Mott-Hubbard insulators, the lowest optical ``interband'' excitation creates
an empty site and a doubly occupied site, i.e., a hole in the LHB and a particle in the UHB.\@
A Hubbard exciton can be regarded as a bound state of an empty site and a doubly occupied site,
moving in a background of singly occupied sites.
Studies of excitons in correlated electron systems thus far have focused on one- (1D) or two-dimensional (2D)
systems. Remarkably, it has been found that exciton binding can be driven by either the Coulomb energy
or the kinetic energy.
The former is found in the 1D extended Hubbard model, which takes into account
the Coulomb interaction $V$ between nearest or next-nearest neighbor
sites.\cite{vandenbrink95a,gallagher97,neudert98,essler01,huebsch01,jeckel03,moskvin03,kim04,matsueda05}
Since {\it both} the Mott-Hubbard gap and the attractive interaction for exciton binding result
from Coulomb interactions, one expects different physics compared to band insulators.
In fact, excitons are only formed below the gap if $V$ exceeds a critical
value.\cite{vandenbrink95a,gallagher97,essler01,jeckel03}
For smaller values of $V$, an excitonic resonance is found in the continuum above the gap,
strongly affecting the line shape of the optical conductivity $\sigma(\omega)$.\cite{essler01,jeckel03,matsueda05}
An exciton below the gap has been observed in 1D Ni-halogen chains,\cite{ono04,ono05} and this
exciton contributes to the gigantic non-linear optical response observed in these
compounds.\cite{kishida,ono04,ono05,matsueda05}

The kinetic energy is of prime importance for excitons in the 2D
cuprates,\cite{clarke,wang,zhangng,wrobel,kuzian,simon,hanamura,moskvin,gomi,itoh,collart,ellis}
which are of charge-transfer type.
The dispersion of a spinless charge-transfer exciton is of order $t$, {\it larger} than the
single-particle dispersion, which is suppressed to $\sim J$ by antiferromagnetic (AF) correlations.
Thus exciton formation reduces the {\it kinetic} energy, \cite{clarke,wang,zhangng,wrobel,kuzian}
which bears resemblance to a possible mechanism for Cooper pair formation in high-$T_c$
superconductors.\cite{wrobel,hirsch,molegraaf02a}
Experimentally, the exciton dispersion has been studied by electron-energy-loss spectroscopy\cite{wang}
and by resonant inelastic x-ray scattering (RIXS).\cite{collart,ellis}
It has been claimed that the dispersion is indeed large,\cite{wang,collart} but recent high-resolution
RIXS data\cite{ellis} indicate that the exciton dispersion is suppressed by the coupling to phonons.

Here, we report on the observation of an excitonic resonance in the optical conductivity $\sigma(\omega)$
of the 3D Mott-Hubbard insulators YTiO$_3$ and SmTiO$_3$.
The former is ferromagnetic below $T_c = 27$\,K, the latter antiferromagnetic below $T_N = 53$\,K,
both exhibit orbital order.\cite{akimitsu01a,iga04a,komarek}
Due to the orbital multiplicity in these $d^1$ spin $S=1/2$ compounds, the upper Hubbard band
consists of a series of different $d^2$ multiplets.
In YTiO$_3$, the lowest multiplet is identified with a peak at 2.55\,eV, whereas a strongly
temperature-dependent peak at 1.95\,eV is
attributed to an excitonic resonance. For a proper determination of $U$ it is essential to
take excitonic effects into account.
We discuss the possible relevance of the kinetic energy for exciton
formation in {\em orbitally} ordered compounds, similar to the case of a 2D antiferromagnet.
Our results provide the experimental basis to disentangle the role of Coulomb and kinetic
energy in 3D Mott-Hubbard insulators.

The spectral weight of the LHB-UHB excitation is expected to depend
on the nearest-neighbor spin-spin correlations.\cite{kovaleva04a,oles05a,lee05a,goessling08}
In a single-band model, the spectral weight vanishes in the case of ferromagnetic order due to
the Pauli principle, and one expects a strong change of the spectral weight as a function of the
temperature $T$ at the magnetic ordering transition. In a multi-orbital system, the spectral
weight also depends on the orbital occupation. This kind of analysis has been applied to a
number of compounds with different transition-metal ions (Mn, V, Ru,
Mo).\cite{khaliullinrev,ahn00,tobe01a,kovaleva04a,lee05a,lee02a,kim04a,rauer,miyasaka02,tsvetkov04a,khaliullin04a,oles05a,goessling08}
For instance in LaMnO$_3$ and LaSrMnO$_4$, a quantitative description of the experimentally
observed $T$ dependence of the spectral weight has been obtained.\cite{kovaleva04a,goessling08}
In the manganites, the $T$ dependence is entirely ascribed to the spin-spin correlations,
whereas the orbital occupation is assumed to be independent of $T$. This reflects the large
ligand-field splitting $\Delta_{eg}$ of roughly 1\,eV of the $e_g$ orbitals in these
$d^4$ compounds.\cite{kovaleva04a,goessling08}
Here, we show that the $T$ dependence of the spectral weight of YTiO$_3$ and SmTiO$_3$ is {\it not}
dominated by the spin-spin correlations. This is particularly evident for SmTiO$_3$, where the sign
of the $T$ dependence of the spectral weight depends on the polarization. This behavior can be
attributed to small changes of the orbital occupation in these $t_{2g}$ compounds.

The paper is organized as follows. Section II addresses the experimental details.
The optical conductivity of YTiO$_3$ and SmTiO$_3$ is reported in section III.\@
In section III.A we first discuss the multiplet assignment and argue that the lowest peak has to be
interpreted as an excitonic resonance in both compounds. A possible contribution of the kinetic
energy to exciton binding in the case of antiferro-{\it orbital} order is proposed in section III.B.\@
In section III.C we discuss the temperature dependence of the spectral weight and the relevance of
spin-spin correlations and orbital occupation. The anisotropy of the spectral weight of the lowest
multiplet in YTiO$_3$ is addressed in section III.D.\@ A summary and conclusions are given in
section IV.\@ The role of oxygen defects for the analysis of ellipsometric data of YTiO$_3$
is discussed in the appendix.

\section{Experimental}

Single crystals of YTiO$_3$ and SmTiO$_3$ were grown using the floating-zone technique.
The crystal quality and stoichiometry were checked by x-ray diffraction, EDX, and polarization
microscopy. The crystals are single phase and single domain.
From magnetization measurements (SQUID, PPMS) we find that YTiO$_3$ becomes ferromagnetic below
$T_c$=27\,K and SmTiO$_3$ antiferromagnetic below $T_N$=53\,K.\@
Further details on crystal preparation and characterization can be found in Ref.~\onlinecite{RothPhD}.\@
In YTiO$_3$, four-sublattice orbital order has been reported up to
room temperature.\cite{akimitsu01a,iga04a}
In both compounds an orbital-ordering transition has not been observed, i.e., they are considered
to be orbitally ordered up to the melting temperature, or, in other words, the distortions arising
from the orbital occupation do not break the crystal symmetry.

Generalized ellipsometric data\cite{schubert96a} was obtained using
a rotating-analyzer ellipsometer (Woollam VASE) equipped with a retarder between polarizer
and sample. The angle of incidence was $70^{\circ}$.
Immediately after polishing, the sample was kept in an UHV cryostat. The measurement
background pressure of $p < 10^{-9}$\,mbar has been achieved by a bakeout at 400\,K for 24\,h.
Window effects have been corrected using a standard Si wafer. In
orthorhombic $R$TiO$_3$, only the diagonal elements $\sigma^{a}$,
$\sigma^{b}$ and $\sigma^{c}$ of the complex optical conductivity
tensor $\sigma(\omega) = \sigma_1 + i \sigma_2$ are finite.
In YTiO$_3$, we have determined $\sigma(\omega)$ from
the normalized M\"{u}ller matrix elements $m^i_{12}$, $m^i_{21}$, $m^i_{33}$, and $m^i_{34}$,
where $i = 1$ -- 4 denotes different orientations of the sample,
namely with $s$-polarized light parallel to the crystallographic $a$ and $b$ ($a^*$ and $c$)
axes on the $ab$ ($a^*c$) surface, where $a^*=[110]$ within the $Pbnm$ space
group. In SmTiO$_3$, $\sigma(\omega)$ has been determined from measurements on $bc$ and $ab$
surfaces.

\begin{figure}[t]
  \includegraphics[width=0.9\columnwidth,clip]{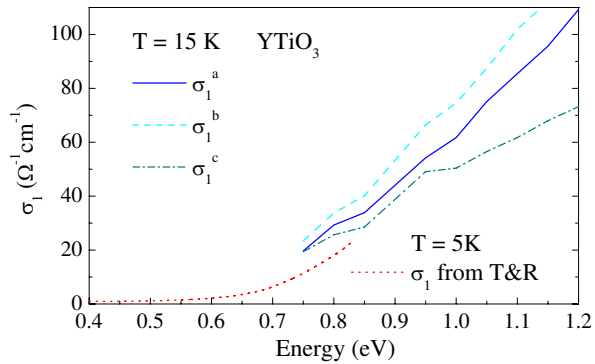}
  \caption{(Color online) Optical conductivity of YTiO$_3$ in the vicinity of the onset of excitations
  across the gap. Good agreement is observed between our ellipsometry data and results determined
  from the combination of transmittance and reflectance measurements.\cite{rueckamp05a}
  }
\label{fig:compareRueckamp}
\end{figure}

Ellipsometry is a surface sensitive technique, thus one has to consider the possible contribution
of surface contaminations or adsorbate layers. To this end we have polished and measured a sample
of YTiO$_3$ several times, both in UHV and under ambient conditions, and for different angles of
incidence. The raw data show small variations which are attributed to the surface.
A consistent description of all data sets for the two distinct surface orientations has been
achieved by assuming a non-absorbing cover layer, where only the thickness $d \leq 2 $\,nm
of this layer has been allowed to vary for different data sets. For an extensive discussion
of the data analysis, we refer to Ref.\ \onlinecite{goesslingthesis}. The particular choice
of the cover layer has a certain influence on the absolute value of $\sigma(\omega)$, but
we emphasize that the temperature dependence is hardly affected.
We also have checked carefully that the observed temperature dependence reflects the
properties of YTiO$_3$ and is not caused by changes of the cover layer, i.e.,
adsorbates.\cite{goesslingthesis} We observed changes of the cover layer if we start with a
base pressure of $p = 10^{-7}$\,mbar, but not for $p < 10^{-9}$\,mbar.
In Fig.\ \ref{fig:Hubbard} we plot $\sigma_1^{a}$, $\sigma_1^{b}$, and $\sigma_1^{c}$ of YTiO$_3$
from 0.75 to 5.8\,eV at 15\,K.\@
The data are consistent with the unpolarized room-temperature data of Ref.\ \onlinecite{arima}
and with infrared transmittance and reflectivity results obtained in our group.\cite{rueckamp05a}
The latter revealed an onset of interband excitations at about 0.6\,eV (see Fig.\ \ref{fig:compareRueckamp}).
Recently, the effect of oxygen defects at the surface of YTiO$_3$ has been discussed.\cite{kovaleva07}
We address this issue in the appendix.

\section{Results}

Undoped YTiO$_3$ and SmTiO$_3$ are Mott-Hubbard insulators. In the ground state there is a single electron
in the $3d$ shell at each Ti site. It is well accepted that the absorption above the gap corresponds to
excitations from the LHB to the UHB, i.e., to the creation of an empty and a doubly occupied site,
$|d^1d^1\rangle \rightarrow |d^0 d^2\rangle$. The strong increase of $\sigma_1(\omega)$ above
$\approx$\,4.5\,eV (see Fig.\ \ref{fig:Hubbard}) reflects the onset of charge-transfer excitations from
the O$_{2p}$ band to the UHB, $|d^1p^6\rangle \rightarrow |d^2p^5 \rangle$.
The difference in spectral weight can be attributed to the Ti-O hopping $t_{pd}$:
$\sigma_1(\omega) \propto t_{pd}^2$ for charge-transfer excitations and
$\sigma_1(\omega) \propto t_{pd}^4/\Delta^2$
for Mott-Hubbard excitations, where $\Delta$ denotes the charge-transfer energy.

For YTiO$_3$, photoemission and inverse photoemission spectroscopy\cite{fujimori92a,bocquet96a,morikawa96a,RothPhD}
yield $\Delta\cong6$\,eV and an on-site Coulomb interaction $U \! \approx \! 5$\,eV,\cite{Uav}
where $U$ denotes the Coulomb repulsion if both electrons occupy the {\em same} real orbital.
In a single-band Hubbard model, the splitting between LHB and UHB is given by $U$
(cf.\ inset of Fig.\ \ref{fig:Hubbard}). However, for a quantitative description
of $\sigma(\omega)$ and for a reliable peak assignment one has to take
all five $3d$ orbitals into account.\cite{kovaleva04a,oles05a,lee05a,goessling08}

\subsection{Multiplet assignment and Hubbard exciton}

\begin{figure}[t]
  \includegraphics[width=0.9\columnwidth,clip]{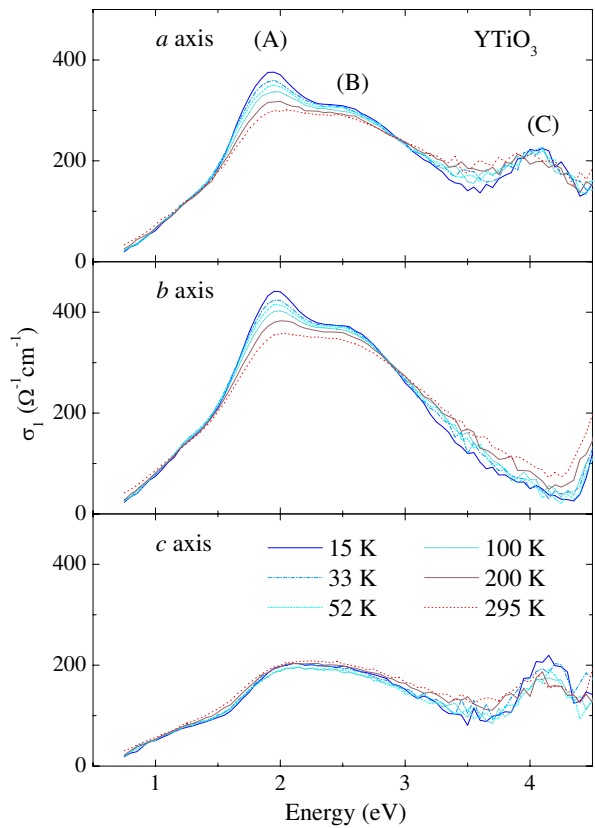}
  \caption{(Color online) Optical conductivity of YTiO$_3$ below the onset of
  charge-transfer excitations,
  i.e., in the range of the lowest excitations from the lower to the upper Hubbard band.
  Peak B is attributed to the lowest multiplet, whereas peak A is identified as an excitonic
  resonance. Peak C reflects the lowest excitation to an $e_g$ orbital.
   }
\label{fig:Ysig}
\end{figure}

\begin{figure}[t]
  \includegraphics[width=0.9\columnwidth,clip]{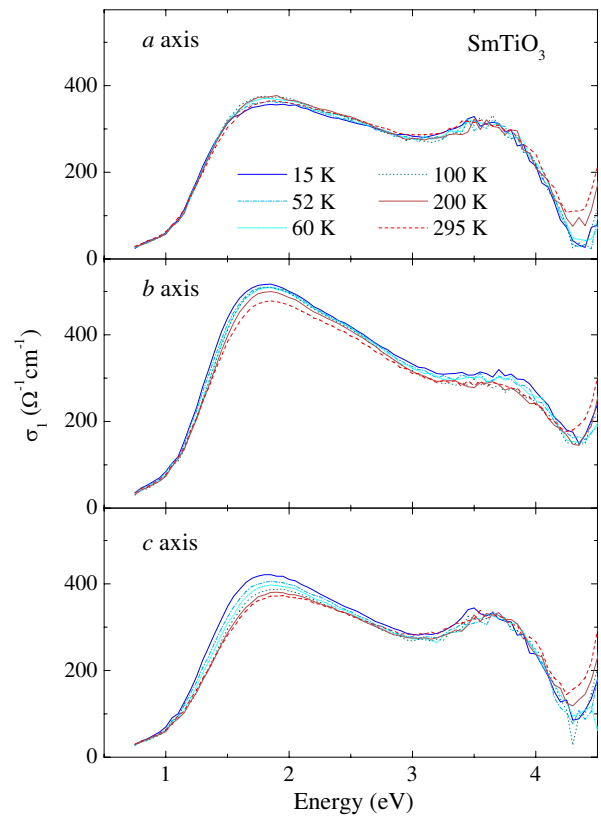}
  \caption{(Color online) Optical conductivity of SmTiO$_3$.
   }
\label{fig:Smsig}
\end{figure}

Figures \ref{fig:Ysig} and \ref{fig:Smsig} focus on the inter-Hubbard-band excitations of
YTiO$_3$ and SmTiO$_3$ below 4.5\,eV.\@
In YTiO$_3$, three peaks are observed at 1.95 (A), 2.55 (B), and 4.15\,eV (C).
In SmTiO$_3$, we find two pronounced peaks at 1.9 and 3.7\,eV.\@ Additionally,
there is a shallow shoulder at 2.5\,eV, particularly noticeable for the $b$ axis.


For a Mott-Hubbard insulator, one expects that a local multiplet calculation
yields a reasonable assignment of the LHB-UHB excitations.\cite{kovaleva04a,oles05a,lee05a,goessling08}
In terms of local multiplets,
the excited $|d^0d^2\rangle$ states can be distinguished according to the $d^2$ sector, because
$d^0$ is an empty shell. The $d^2$ sector is split into a series of
multiplets by the electron-electron interaction, the crystal field,
and the hybridization with the ligands.\cite{tanabesugano} We start
from cubic symmetry, in which case the crystal field and the
hybridization give rise to a splitting of the $3d$ orbitals into a
triply degenerate $t_{2g}$ level and a doubly degenerate $e_g$ level
at higher energy. The splitting is denoted by 10\,Dq, which roughly
can be estimated as $2\pm0.5$\,eV [\onlinecite{higuchi,ulrich08,iga04a,schmitz05c}].
The electron-electron interaction within the $3d$ shell can be parameterized by the three
Slater integrals $F^0$, $F^2$ and $F^4$.
Values of $F^2$=6.75\,eV and $F^4/F^2\approx 5/8$ are characteristic for $d^2$ Ti$^{2+}$ ions
in a crystal.\cite{tanabesugano}
The only parameter that can be adapted is $F^0$, which drastically deviates in a solid from the
ionic value due to screening effects.

For $F^0=3.60$\,eV (or $U \! \approx \! 4.5$\,eV [\onlinecite{Uav}])
the $|d^1d^1\rangle \rightarrow |d^0 d^2\rangle$ excitation energies are given
in Fig.\ \ref{fig:levels},
focusing on the four multiplets lowest in energy: the triplet
$^3T_{1}$, the singlets $^1T_{2}$ and $^1E$, and the triplet
$^3T_{2}$. For an intuitive picture we consider the strong crystal-field
limit (10\,Dq $\gg U$), as sketched on the right hand side of Fig.\ \ref{fig:levels}.
In this limit, there is one electron in the $t_{2g}$ level and one in the $e_g$ level
in the $^3T_2$ state, whereas both electrons occupy the $t_{2g}$ level in the three other
states.
It is common to consider the simplified Kanamori scheme\cite{lee05a} with the Hund on-site exchange
coupling $J_H = \frac{2.5}{49}F^2 + \frac{22.5}{441} F^4$, resulting in $J_H \! = \! 0.6 \pm 0.1$\,eV
for $d^2$ Ti$^{2+}$.
For $U \! \cong \! 4-5$\,eV, the Kanamori scheme predicts the lowest excitation into the $^3T_1$
triplet at $U-3J_H\approx 2-3$\,eV, separated from the singlets $^1T_{2}$ and $^1E$ by
$2J_H \approx 1.2$\,eV (reflecting Hund's rule) and from the $^3T_2$
state by 10\,Dq\,$\approx 2$\,eV, in qualitative agreement with the
result of the rigorous calculation shown in Fig.\ \ref{fig:levels}.

\subsubsection{YTiO$_3$}

\begin{figure}[tb]
     \includegraphics[width=0.90\columnwidth,clip]{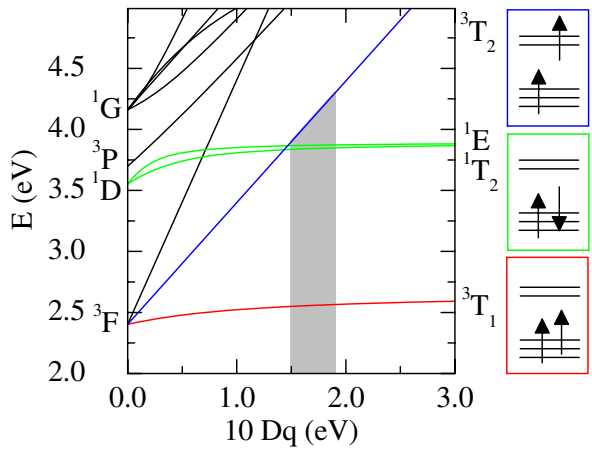}
    \caption{(Color online) Left: Calculated energies for a $|d^1d^1\rangle$ $\rightarrow |d^0 d^2\rangle$ excitation with
    different $d^2$ final states in a cubic crystal-field.\cite{tanabesugano}
    The Slater integrals were chosen as $F^0$ = 3.60\,eV, $F^2$=6.75\,eV, and $F^4$=4.55\,eV,
    corresponding to $U\! = \!  4.5$\,eV [\onlinecite{Uav}] and $J_H  \approx 0.6$\,eV.\@
    For 10\,Dq = 0 the ionic multiplet structure is obtained.
    For 10\,Dq $\approx$ 1.5 - 1.9\,eV (grey area) the energy of 4.15\,eV for peak C is well described
    by excitations into the $^3T_2$ state.
    Right: sketch of the orbital occupation in the strong crystal-field limit
    for the $^3T_1$ triplet (red), the $^1T_2$ and $^1E$ singlets (green),
    and the $^3T_2$ triplet (blue).
    }
    \label{fig:levels}
\end{figure}

Figure \ref{fig:levels}  clearly shows that the $^3T_1$ state is the lowest multiplet, more than
1.2\,eV below the next multiplet for any reasonable choice of 10\,Dq. Thus the small splitting
of 0.6\,eV between peaks A and B in YTiO$_3$ cannot be identified with the difference between
the $^3T_1$ state and any other multiplet. We conclude that both peaks A and B are related to
excitations into the $^3T_1$ state. Peak C can be attributed to the
$^3T_2$ state, since only excitations into triplet states are
allowed from a fully polarized ferromagnetic ground state within an
electric dipole approximation. Excitations to the singlet states
$^1T_2$ and $^1E$ require a spin flip and thus are suppressed, at
least at low temperatures.

In the following, we discuss three
scenarios for the splitting between peaks A and B: deviations from
cubic symmetry, band structure effects, and an excitonic resonance.
The deviation from cubic symmetry lifts the degeneracy of the $t_{2g}$ orbitals and thereby also
of the $^3T_1$ state. The $t_{2g}$ splitting was found to be $\approx $\,0.25\,eV in infrared
transmittance,\cite{rueckamp05a} Raman scattering, \cite{ulrich06} and RIXS measurements.\cite{ulrich08}
This is clearly too small to explain the splitting between peaks A and B.\cite{t2gsplitting}

Now we address the possible role of band structure effects. Based on the actual crystal structure,
a LDA+DMFT study of YTiO$_3$ by Pavarini {\it et al.}\cite{pavarini05a} does not show a splitting
of the lowest peak in $\sigma_1(\omega)$.
For $U$=5\,eV and $J_H$=0.64\,eV, this peak has been predicted at 3.3\,eV, and the optical gap
is expected roughly at 1.5\,eV.\@ This large value of the gap suggests that a smaller value of $U$
is more appropriate. Good agreement between the prediction for the lowest peak and the observed
energy of 2.55\,eV of peak B can be obtained by assuming $U \! \approx \! 4.3$\,eV.\@
This value of $U$ also yields a good description of the optical gap. Moreover, it corroborates the
validity of our local multiplet calculation discussed above (see Fig.\ \ref{fig:levels}), which
for $U$=4.5\,eV predicts the lowest peak at about 2.5\,eV.\@
We stress that it is unreasonable to identify peak A at 1.95\,eV with the peak found in LDA+DMFT,
since this would require to assume a still smaller value of $U$, resulting in a very small gap.
Indeed the LDA+DMFT calculation finds a metallic state for $U$=3.5\,eV.\cite{pavarini05a}

Also the LDA+DMFT study by Craco {\it et al.}\cite{craco07} for the ferromagnetic phase of YTiO$_3$
finds a single peak in $\sigma_1(\omega)$. Based on the parameter values of $U \! = \!4.75$\,eV and
$J_H \! = \!1.0$\,eV, Craco {\it et al.\ } attribute this peak to peak A observed at 1.95\,eV in our
data. However, $J_H$ is not expected to deviate strongly from the ionic value of $J_H \! = \!0.6 \pm 0.1$\,eV
discussed above. The lowest peak in $\sigma_1(\omega)$ is located at about $U-3J_H$, thus the choice of
$J_H \! = \!1.0$\,eV strongly underestimates the peak frequency. We emphasize that both LDA+DMFT
studies\cite{pavarini05a,craco07} find a {\it single} peak in $\sigma_1(\omega)$. Both studies investigate
an effective Hamiltonian for the $t_{2g}$ sector, i.e., excitations to the higher-lying $^3T_2$ multiplet
(peak C) are not considered.

Experimental data neither support a splitting due to band-structure effects.
In photoemission (PES) data of YTiO$_3$ the LHB is a single peak $\approx$\,1.3\,eV
below the Fermi level.\cite{fujimori92a,morikawa96a,RothPhD,arita} In inverse PES on
Y$_{1-x}$Ca$_x$TiO$_3$ ($x$=0 [\onlinecite{arita}] and 0.4 - 0.8 [\onlinecite{morikawa96a}])
the UHB can be identified with the lowest peak or shoulder $\approx$ 1.5 - 2\,eV above the Fermi
level. Both PES and inverse PES agree with the LDA+DMFT result\cite{pavarini05a} for
$U\! = \!4$ - 5\,eV.\@ Finally, $U \approx 5.3$\,eV has been derived from $2p$ core-level
PES [\onlinecite{bocquet96a}] (see [\onlinecite{Uav}] for the comparison of parameters derived
by different methods).

Altogether, both theoretical and experimental results support our interpretation that the splitting
of 0.6\,eV between peaks A and B does not result from the band structure and that peak B
at 2.55\,eV is the dominant excitation.

In contrast to (inverse) PES, the optical conductivity reflects
particle-hole excitations and thus is sensitive to interactions
between the particle in the UHB (i.e.\ a double occupancy) and the
hole in the LHB.\@ These interactions are also neglected in the
LDA+DMFT calculations\cite{pavarini05a,craco07} of $\sigma_1(\omega)$.
We therefore identify peak B at 2.55\,eV as a particle-hole excitation
in which the particle and the hole are well separated, whereas peak A
at 1.95\,eV is interpreted as an excitonic resonance,
where the particle and the hole remain close to each other.
Note that peak A does not lie below the gap, i.e., it is not a truly
bound exciton, but a resonance within the continuum. As discussed above,
a Hubbard exciton may arise due to the
attractive Coulomb interaction between the particle and the hole.
The nearest-neighbor electron-electron repulsion $V$ of the extended Hubbard
model\cite{vandenbrink95a,gallagher97,neudert98,essler01,huebsch01,jeckel03,moskvin03,kim04,matsueda05}
is equivalent to a particle-hole
attraction $-V$ [\onlinecite{magnon}]. We are not aware of an accurate
experimental value of $V$ for the titanates, but it is reasonable to assume
$V \! \leq \! 1$\,eV.\@ For the 1D charge-transfer insulator SrCuO$_2$,
a value of $V \approx 0.6$\,eV has been derived from the comparison of the
line shape of the excitonic resonance in $\sigma_1(\omega)$ with predictions
from dynamical density-matrix renormalization group calculations for an
effective extended Hubbard model.\cite{kim04}
More detailed theoretical studies of the extended Hubbard model in 3D are required
to decide whether the nearest-neighbor Coulomb interaction is sufficient to explain
the splitting of 0.6\,eV observed between peaks A and B.

\subsubsection{SmTiO$_3$}

The magnetic ground state of $R$TiO$_3$ changes from ferromagnetic to antiferromagnetic
as a function of the size of the $R$ ions.\cite{mochizuki04a,mochizuki04b}
This change is accompanied by a crossover of both the character of the distortions of
the oxygen octahedra and of the orbital-ordering pattern.\cite{mochizuki04a,mochizuki04b,kubota,komarek}
The optical conductivity of the antiferromagnet SmTiO$_3$ is given in Fig.\ \ref{fig:Smsig},
focusing on the range of the Mott-Hubbard bands below the onset of charge-transfer excitations
at about 4.5\,eV.\@
At 300\,K we observe two pronounced peaks at 1.9 and 3.7\,eV and a shallow shoulder
at 2.5\,eV, most evident for the $b$ axis.
The multiplet structure discussed above for YTiO$_3$ also applies to SmTiO$_3$.
In particular, one does not expect appreciable changes of the Slater integrals, i.e., of the
electronic parameters $U$ and $J_H$. This is corroborated by the LDA+DMFT calculations by
Pavarini {\it et al.},\cite{pavarini05a} which predict the same peak frequency for $\sigma_1(\omega)$
in YTiO$_3$ and in the antiferromagnet LaTiO$_3$. Therefore we identify the shoulder at 2.5\,eV
in SmTiO$_3$ with peak B at 2.55\,eV in YTiO$_3$, whereas the asymmetric peak at 1.8\,eV is
attributed to an excitonic resonance (see below), equivalent to peak A in YTiO$_3$.

In contrast to the peak energies, the band width or equivalently the hopping matrix elements
are expected to change significantly, resulting from the different Ti-O-Ti bond angles.
A larger band width of SmTiO$_3$ agrees with the observation from transmittance measurements\cite{rueckamp05a}
that the gap in SmTiO$_3$ is about 0.2\,eV lower than in YTiO$_3$.
Additionally, an increase of the hopping amplitudes gives rise to an increase of the spectral
weight. This is further enhanced by the change of the orbital ground state.\cite{mochizuki04a,mochizuki04b}
Experimentally, we find an increase of $N_{\rm eff}$ from YTiO$_3$ to SmTiO$_3$ of
roughly 25\%, 50\%, and 100\% for the $a$, $b$, and $c$ axes, respectively.

\begin{figure}[t]
  \includegraphics[width=0.95\columnwidth,clip]{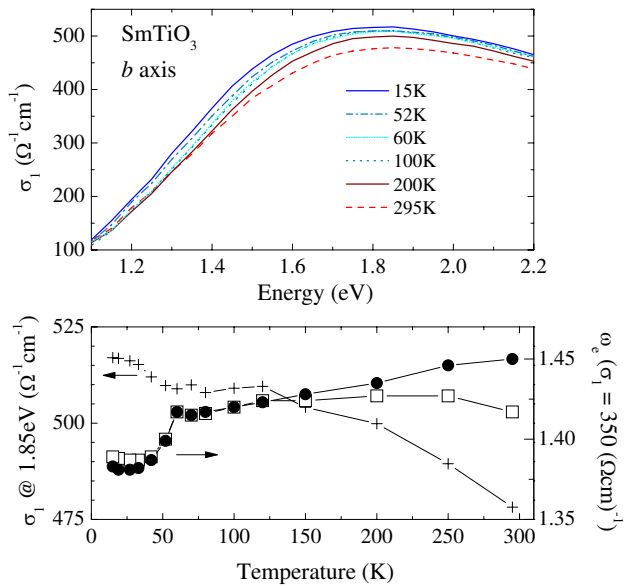}
  \caption{(Color online) Top: Leading edge of the optical conductivity $\sigma_1^b(\omega)$ of SmTiO$_3$.
  Bottom: Temperature dependence of $\sigma_1^b(\omega=1.85$\,eV$)$ (crosses, left axis) and of the
  leading edge $\omega_e$, which we define by $\sigma_1^b(\omega_e) = 350$\,($\Omega$cm)$^{-1}$ (full symbols,
  right axis). Open symbols show $\tilde{\omega}_e$ corrected for the change of the spectral weight, i.e.,
   $\sigma_1^b(\tilde{\omega}_e) = c\cdot 350$\,($\Omega$cm)$^{-1}$ with
   $c = \sigma_1^b(1.85\,$eV$,T) / \sigma_1^b(1.85\,$eV$,60$\,K$)$. }
\label{fig:Smedge}
\end{figure}

As discussed above for YTiO$_3$, an interpretation of the peak at 1.9\,eV in terms of the lowest multiplet
is hard to reconcile with the LDA+DMFT result,\cite{pavarini05a} unless excitonic effects are considered.
An exciton interpretation is supported by the temperature dependence of the peak frequency observed for
the $b$ and $c$ axes, showing an anomalous softening with decreasing temperature and an anomaly at $T_N$.
For the $b$ axis we demonstrate this softening in Fig.\ \ref{fig:Smedge}.
We focus on the frequency $\omega_e$ of the leading edge, which we define as
$\sigma_1^b(\omega_e)= 350$\,($\Omega$cm)$^{-1}$. This has the advantage that $\omega_e$ and also
its temperature dependence can be determined more accurately than the peak frequency itself.
The disadvantage is that a softening of $\omega_e$ in principle can be caused not only by a softening
of the peak frequency, but also by an increase of either the spectral weight or the line width,
and by a change of the line shape.
We find a jump-like decrease of $\omega_e$ at $T_N$, see lower panel of Fig.\ \ref{fig:Smedge}.
This cannot be attributed to an increased line width, since the thermal contribution to the line width
is expected to decrease with decreasing temperature. Moreover, for {\it spin-carrying} particles one
expects that the band width is reduced upon entering the AF ordered state, thus the gap is expected
to {\it harden}.
For an estimate of the $T$ dependence of the spectral weight we consider the value of $\sigma_1(\omega)$
at the peak frequency. We find an increase of $\sigma_1^b(\omega$=1.85\,eV) upon cooling below
$T_N$ (see crosses in bottom panel of Fig.\ \ref{fig:Smedge}), indicating an increase of the spectral weight.
We use this $T$ dependence of the absolute value to determine a corrected frequency of the leading
edge, $\tilde{\omega}_e$, defined as $\sigma_1^b(\tilde{\omega}_e) = c\cdot 350$\,($\Omega$cm)$^{-1}$
with $c = \sigma_1^b(1.85\,$eV$,T) / \sigma_1^b(1.85\,$eV$,60$\,K$)$ (open symbols in Fig.\ \ref{fig:Smedge}).
This shows that the shift of $\omega_e$ is not due to the change of the spectral weight,
it is caused mainly by a softening of the peak frequency or
a change of the line shape. Both can be rationalized by the attractive interactions
responsible for exciton formation, pulling the spectral weight to lower frequencies.

The excitonic binding due to the kinetic energy is enhanced in the AF ordered state, as discussed for
2D compounds in the introduction. Remarkably, the peak frequency of 1.95\,eV is independent of temperature
in the ferromagnet YTiO$_3$.
It is an interesting question whether this difference between the two compounds arises from a change of
the screening of nearest-neighbor Coulomb interactions, of the magnetic ground state, or of the orbital
ordering pattern.
A decisive identification of the driving force for exciton formation requires further theoretical
investigations of the extended multi-orbital Hubbard model in 3D.\@

The peak at 3.7\,eV coincides with the minimum of $\sigma_1(\omega)$ observed in YTiO$_3$. This peak can
be attributed to the lowest singlet multiplet ($^1T_2$ and $^1E$ in cubic symmetry). Due to the spin selection
rule, the excitation to the singlet state is suppressed in ferromagnetic YTiO$_3$, but it is allowed in
antiferromagnetic SmTiO$_3$.
This feature is expected at about $2J_H \approx 1.2$ -- 1.3\,eV above the lowest triplet peak, providing
further support for the assignment of peak B at 2.5\,eV and the excitonic character of the peaks at 1.8\,eV
in SmTiO$_3$ and 1.95\,eV in YTiO$_3$.

\subsection{Hubbard exciton and orbital order}

For a 2D Mott-Hubbard insulator with AF exchange $J$ on a square lattice, exciton formation
is governed by the {\em kinetic} energy.\cite{clarke,wang,zhangng,wrobel,kuzian}
The motion of a single particle is hindered by the interaction of its spin with the AF background.
This can be described in terms of a spin polaron. Hopping of the bare particle on the energy scale $t$
results in a trace of misaligned spins. Coherent motion of the dressed polaronic quasiparticle requires
the emission of magnons, i.e., the bare band width $\sim$\;$t$ is reduced to the polaronic band width
$\sim$\;$J$, which corresponds to an increase of kinetic energy.
In this case, the kinetic energy is lowered by the formation of {\it spinless} excitons,
which recover a larger band width.

\begin{figure}[tbp]
     \includegraphics[width=0.7\columnwidth,clip]{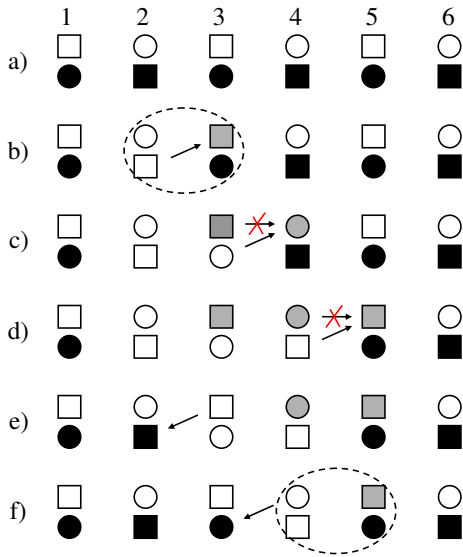}
    \caption{
    Sketch of the suggested formation and propagation of a Hubbard exciton (dashed line).
    We consider two types of orbitals (circles and squares, e.g., $d_{xy}$ and $d_{xz}$)
    per site, where hopping is zero between orbitals of different type (crossed out arrows).
    Full (open) symbols denote occupied (empty) orbitals.
    (a) Ground state with antiferro-orbital order.
    (b) Creation of a hole and a double occupancy on sites 2 and 3, respectively.
    (c)--(f) Propagation of the double occupancy, the hole, or of an exciton (see main text for
    more details).
    }
    \label{fig:kinetic}
\end{figure}

This mechanism may contribute to exciton binding in antiferromagnetic SmTiO$_3$, but not in
ferromagnetic YTiO$_3$.
It is promising to investigate whether a similar mechanism is at work in the case of antiferro-{\em orbital}
order. For illustration and simplicity, we consider a 1D model with two orbitals per site,
e.g., $d_{xy}$ and $d_{xz}$ for a chain running along the $x$ direction. In Fig.\ \ref{fig:kinetic},
the two types of orbitals are denoted by circles and squares, respectively.
Hopping between neighboring sites is allowed only between orbitals of the same type;
it is zero between orbitals of different type.
Black and grey symbols in Fig.\ \ref{fig:kinetic} refer to occupied orbitals, whereas empty symbols denote
empty orbitals. The ground state in Fig.\ \ref{fig:kinetic}a exhibits antiferro-orbital order, i.e.,
$xy$ (circles) and $xz$ orbitals (squares) are occupied in an alternate fashion.
The empty orbitals are at higher energies due to, e.g., the ligand-field splitting.
An excitation from the LHB to the UHB, i.e., $|d^1d^1\rangle \rightarrow |d^0 d^2\rangle$,
is illustrated in Fig.\ \ref{fig:kinetic}b. Site 2 is empty, and site 3 is doubly occupied.
The motion of the double occupancy to sites 4 and 5 is depicted in Figs.\ \ref{fig:kinetic}c
and d, respectively.
The central point is that this motion leaves a trace of orbitally excited states, i.e.,
on sites 3 and 4 the energetically unfavorable orbitals are occupied (grey symbols).
This results from the restriction that hopping is only allowed within the same type of
orbital. As discussed above for the case of spins, coherent motion of the quasi-particle requires
the emission of orbital excitations, in our example the de-excitation of sites 3 and 4.
Therefore, the band width is reduced from the bare band width $\sim t$ to the energy scale
of the orbital excitations, corresponding to an increase of kinetic energy.
However, if the hole accompanies the double occupancy forming an exciton (dashed line), the motion
of the hole heals out the trace of excited orbitals, see Fig.\ \ref{fig:kinetic}e and f.
Therefore, the motion of the
exciton is not hindered by the antiferro-orbital order, and the exciton can hop on a larger
energy scale than the hole or the double occupancy individually. Thus exciton formation here
is equivalent to a {\it gain of kinetic energy}.

More detailed knowledge on the value of the nearest-neighbor Coulomb interaction $V$
and its relationship to the binding energy in 3D Mott-Hubbard insulators is required to decide
whether this mechanism is realized in YTiO$_3$. The orbital ordering pattern in YTiO$_3$
is more complex than simple antiferro-orbital order,\cite{akimitsu01a,iga04a}
and hopping between orbitals of different type is not exactly zero.
Still Fig.\ \ref{fig:kinetic} may be relevant for the $ab$ plane, since hopping from the
lowest orbital on one site (a ``circle'' in Fig.\ \ref{fig:kinetic}) to the lowest orbital
on a neighboring site (equivalent to a ``square'') is 2-3 times smaller than hopping to the
excited states.\cite{schmitz05c,pavarini05a}

\subsection{Temperature dependence of the spectral weight: spin and orbital selection rules}

The spectral weight is determined by the spin and orbital selection
rules.\cite{khaliullinrev,ahn00,tobe01a,kovaleva04a,lee05a,lee02a,kim04a,rauer,miyasaka02,tsvetkov04a,khaliullin04a,oles05a,goessling08}
Therefore, the $T$ dependence of the spectral weight is expected to reflect changes of the
spin-spin correlations and/or of the orbital occupation. By considering the nearest-neighbor
spin-spin correlations, the absolute value of the spectral weight has been calculated
for instance for LaMnO$_3$ [\onlinecite{kovaleva04a,oles05a,khaliullinrev}] and LaSrMnO$_4$ [\onlinecite{goessling08}].
These calculations yield a convincing description of the experimental results, the maximum
difference is less than a factor of 2.
For a 3D magnet one expects that the spin-spin correlations are small above the ordering temperature.
In fact, the change of the spectral weight above $T_N$ is small in the 3D antiferromagnet
LaMnO$_3$.\cite{kovaleva04a} In contrast, the 2D antiferromagnet
LaSrMnO$_4$ with a N\'{e}el temperature of $T_N = 130$\,K, exhibits a significant $T$ dependence
of the spectral weight up to 300\,K, which can be attributed to enhanced quantum fluctuations
in 2D.\cite{goessling08}

For the lowest excited triplet state ($^3T_1$ in cubic notation) of orbitally ordered YTiO$_3$,
Oles {\it et al.}\cite{oles05a} predicted a change of 25\% of the spectral weight between the
paramagnetic and the ferromagnetic state in the $ab$ and $c$ directions. This can be understood
by the evolution of the nearest-neighbor spin-spin correlation function
$\langle {\bf S}_i \cdot {\bf S}_j + 3/4\rangle$, which equals 1 in the ferromagnetic state
and 3/4 in the paramagnetic state, i.e., a redistribution of 25\%.

\begin{figure}[tbp]
     \includegraphics[width=0.95\columnwidth,clip]{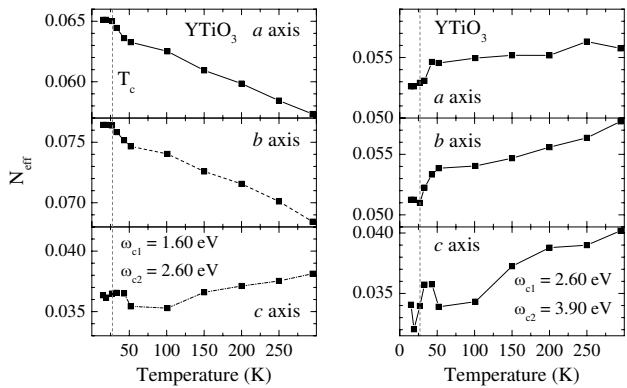}
    \caption{Temperature dependence of the effective carrier concentration $N_{\rm eff}$
    (see Eq.\ \ref{eq:Neff}) of YTiO$_3$
     for $\omega_{\rm c1} \, = \, 1.6$\,eV and $\omega_{\rm c2} \, = \, 2.6$\,eV (left) and
     for $\omega_{\rm c1} \, = \, 2.6$\,eV and $\omega_{\rm c2} \, = \, 3.9$\,eV (right).
    }
    \label{fig:Y_Neff}
\end{figure}

\begin{figure}[t]
  \includegraphics[width=0.95\columnwidth,clip]{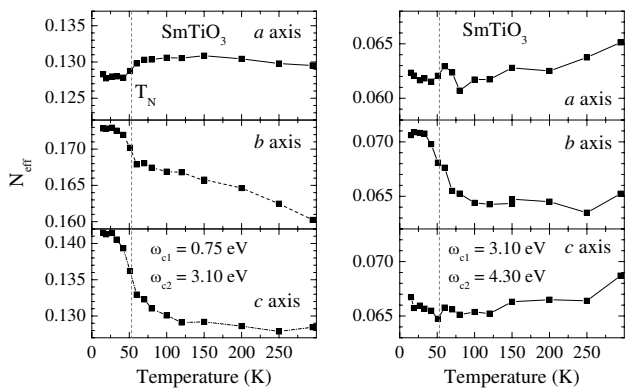}
  \caption{Temperature dependence of the effective carrier concentration $N_{\rm eff}$
    (see Eq.\ \ref{eq:Neff}) of SmTiO$_3$
     for $\omega_{\rm c1} \, = \, 0.75$\,eV and $\omega_{\rm c2} \, = \, 3.1$\,eV (left) and
     for $\omega_{\rm c1} \, = \, 3.1$\,eV and $\omega_{\rm c2} \, = \, 4.3$\,eV (right).   }
\label{fig:SmNeff}
\end{figure}

We analyze the integrated spectral weight in terms of the effective carrier concentration $N_{\rm eff}$,
\begin{equation}
N_{\rm eff} =\frac{2 m V_0}{\pi e^2} \int_{\omega_{\rm c1}}^{\omega_{\rm c2}}\sigma_1(\omega) d\omega
\label{eq:Neff}
\end{equation}
where $\omega_{\rm c1}$ and $\omega_{\rm c2}$ denote the frequency range of interest,
$m$ is the free electron mass, $e$ the elementary charge, and $1/V_0$ the density of Ti ions.
For $V_0$ we use the value observed at 290\,K, which differs from the 2\,K value by less
than 1\%.\cite{komarek} The $T$ dependence of $N_{\rm eff}$ is given in Fig.\ \ref{fig:Y_Neff}
for YTiO$_3$ and in Fig.\ \ref{fig:SmNeff} for SmTiO$_3$.
In YTiO$_3$, the spectral weight increases between 1.6 and 2.6\,eV (left panel of Fig.\ \ref{fig:Y_Neff})
in the $a$ and $b$ directions upon cooling down from room temperature. We find an anomaly in
the vicinity of $T_c$, i.e., an additional increase of spectral weight with decreasing
temperature. This additional increase starts at about 1.5 -- 2 $T_c$ and amounts to less
than 5\% below 50\,K, much smaller than predicted. At the same time, one finds an anomalous
decrease of spectral weight with decreasing temperature between 2.6 and 3.9\,eV, again in
$a$ and $b$. Moreover, the spin selection rule cannot explain that the spectral weight in
this 3D ferromagnet shows a strong $T$ dependence up to 300\,K $>$ 10 $T_c$.


In SmTiO$_3$, we also find pronounced anomalies in the vicinity of the magnetic ordering
temperature, $T_N \, = \, 53$\,K.\@ The strongest change of $N_{\rm eff}$ of up to 10\% is observed
in the $c$ axis, significantly larger than in YTiO$_3$.
In SmTiO$_3$, the {\it sign} of the change at $T_N$ depends on the polarization (see left panel of
Fig.\ \ref{fig:SmNeff}), which certainly cannot be attributed to the spin selection rule in
a $G$-type antiferromagnet. In comparison to the excellent agreement found between experiment and
theory in the manganites,\cite{kovaleva04a,goessling08} this failure appears as a puzzle.

Alternatively, we consider the orbital selection rule.
In the manganites, the orbital occupation has been assumed to be independent of $T$ due to the
large $e_g$ splitting of roughly 1\,eV.\cite{kovaleva04a,goessling08}
In both YTiO$_3$ and SmTiO$_3$, the $t_{2g}$ splitting is only
$\approx $\,0.25\,eV [\onlinecite{rueckamp05a,ulrich06,ulrich08}], opening the possibility for
small changes of the orbital occupation as a function of $T$. A change of the orbital occupation
affects the effective Ti-Ti hopping amplitude $t$, with $N_{\rm eff} \propto t^2 \propto t_{pd}^4/\Delta^2$.
An increase of the occupation of the planar $xy$ orbital may for instance give rise to an increase
of the spectral weight within the $xy$ plane accompanied by a decrease of spectral weight along $z$.
Thus a change of the orbital occupation at $T_N$ can very well account for the observed polarization
dependence.
Based on a detailed analysis of the crystal structure, thermal expansion and magnetostriction of $R$TiO$_3$,
Komarek {\it et al.}\cite{komarek} conclude that magnetism affects the crystal structure, which in turn
drives a change of the orbital occupation.
Remarkably, the shape of the oxygen octahedra changes significantly as a function of temperature,
whereas the variation of the tilt and rotation angles is small.\cite{komarek}
Both the lattice distortions and the orbital occupation adapt in order to enhance the gain of energy
within the spin system. The effect is most pronounced at the magnetic ordering temperature, but extends
also to higher temperatures, in agreement with our data. Moreover, Komarek {\it et al.}\cite{komarek}
pointed out that the change of the orbital occupation is significantly stronger in SmTiO$_3$ than in
YTiO$_3$, again in agreement with our results. The occurrence of pronounced effects in SmTiO$_3$
is attributed to the fact that SmTiO$_3$ is close to the crossover from antiferromagnetic to
ferromagnetic order.\cite{komarek} In the optical data, the effect of the orbital selection rule
possibly overrules that of the spin selection rule, which appears as a failure of the latter.

Note that the change of the lattice constants at $T_N$ can only account for a change
of $N_{\rm eff}$ of the order of 1\%. This estimate is based on the Harrison rules\cite{harrison}
for the hopping amplitudes. We emphasize that binding phenomena such as the formation
of excitons or resonances in general are very sensitive to temperature. With decreasing temperature, the
attractive interactions responsible for the exciton formation pull down the spectral weight to lower
energies, in agreement with the change of the line shape observed in YTiO$_3$ (see Fig.\ \ref{fig:Ysig})
and the shift of the absorption edge of SmTiO$_3$ discussed above (see Fig.\ \ref{fig:Smedge}).

\subsection{Anisotropy}

In order to understand the anisotropy observed in YTiO$_3$ between the $ab$ plane and the $c$ direction,
we address the matrix elements for the optical excitation $|d^1d^1\rangle \rightarrow |d^0 d^2\rangle$.
Our Hamiltonian includes the crystal field, the on-site Coulomb correlations of the $d^2$ configurations,
and the hopping between the two Ti sites (for details, see Ref.\ \onlinecite{schmitz05c}).
We find realistic values of the exchange coupling constants for the different
directions as well as an orbital ground state which is in excellent agreement with x-ray, neutron
and other theoretical results.\cite{iga04a,akimitsu01a,pavarini05a,streltsov05a,rueckamp05a}
From the effective Ti-Ti hopping matrices $t^{ab}$ and $t^{c}$ one can estimate the anisotropy of the
spectral weight from
\begin{equation}
\frac{N^{ab}_{\rm eff}}{N^c_{\rm eff}} =
\frac{N^a_{\rm eff} + N^b_{\rm eff}}{2N^c_{\rm eff}} = \sum_{j=2,3}\frac{(t^{ab}_{1j})^2+(t^{ab}_{j1})^2}{(t^c_{1j})^2+(t^c_{j1})^2} \, ,
\label{aniso}
\end{equation}
where $t_{ij}$ denotes the effective hopping matrix element between the $t_{2g}$ orbitals $i$ and $j$ on adjacent sites.
We obtain $N^{ab}_{\rm eff}/N^c_{\rm eff} \approx 5.1$.
Using the hopping matrices published by other groups, we find a value of
1.1 [\onlinecite{solovyev06}] or 3.5 [\onlinecite{pavarini05a}].
In the next step the optical conductivity has been calculated using the Kubo formula.
We assume a fully polarized ferromagnetic ground state. We address only excitations into the
lowest triplet state with a $t_{2g}^2$ configuration, because here the point-charge approximation
gives reliable results. We predict $N^{ab}_{\rm eff}/N^c_{\rm eff} \approx 4.5$. The small difference
to the value of 5.1 derived from the simplified approach considered in Eq.\ \ref{aniso} arises because
here the energies of the excited states and the Ti-Ti distance are taken into account.
For $\omega_{c1} = 1.6$\,eV and $\omega_{c2} = 2.6$\,eV (see Eq.\ \ref{eq:Neff}), we experimentally find
$N^{ab}_{\rm eff}/N^c_{\rm eff} \approx 2$ (see Fig.\ \ref{fig:Y_Neff}),
within the range predicted by the different theoretical approaches.

\section{Summary and Conclusions}

In summary, we report on optical excitations from the lower to the upper Hubbard band in the
ferromagnet YTiO$_3$ and in the antiferromagnet SmTiO$_3$. At 15\,K we find peaks in the
optical conductivity $\sigma_1(\omega)$ at 1.95, 2.55, and 4.15\,eV in YTiO$_3$ and at
1.8 and 3.7\,eV in SmTiO$_3$, which also exhibits a shallow shoulder at 2.5\,eV.\@
For these Mott-Hubbard insulators, a local multiplet scenario is expected to yield a reasonable peak
assignment, as reported for the manganites.\cite{kovaleva04a,goessling08}
For $U \! \approx \! 4.5$\,eV and $J_H \! = \! 0.6 \pm 0.1$\,eV, our local multiplet calculation
offers a quantitative description of the peak positions at 2.5, 3.7 and 4.15\,eV.\@
The peak at about 2.5\,eV is attributed to excitations into the lowest $d^2$ multiplet ($^3T_1$
in cubic symmetry) with an energy of roughly $U-3J_H$ [\onlinecite{t2gsplitting}].
The peak at 3.7\,eV corresponds to the lowest $d^2$ singlet states, and the peak at 4.15\,eV
is attributed to the lowest state with a $t_{2g}^1 e_g^1$ configuration.
This assignment is in agreement with photoemission and LDA+DMFT results.
The peaks at 1.95\,eV in YTiO$_3$ and 1.8\,eV in SmTiO$_3$ are interpreted in terms of an excitonic
resonance, thereby explaining their low energy.

The temperature dependence of the spectral weight disagrees with predictions based on the spin selection
rule. In YTiO$_3$ the observed temperature dependence is much smaller than predicted, whereas in SmTiO$_3$
even the sign of the temperature dependence disagrees for certain polarization directions, a puzzling result.
However, a small change of the orbital occupation at the magnetic ordering temperature\cite{komarek} can
account for the polarization dependence and also explains the larger temperature dependence found for SmTiO$_3$.
In contrast to the manganites, such a change of the orbital occupation is feasible in $R$TiO$_3$
because the $t_{2g}$ splitting amounts to only 0.25\,eV.\@
Furthermore, the increase of spectral weight at low frequencies with decreasing temperature is in
agreement with an exciton scenario, since binding phenomena are expected to exhibit a strong temperature
dependence. An increased binding at low temperatures pulls down spectral weight to lower frequencies
and also explains the anomalous softening of the leading absorption edge observed in SmTiO$_3$.

The importance of excitonic effects for the description of $\sigma(\omega)$ is well established
for low-dimensional correlated insulators. The attractive interaction responsible for exciton
formation arises from a gain of either Coulomb or kinetic energy.
We have pointed out that exciton formation may lower the kinetic energy in an orbitally ordered state.
Our results call for further theoretical studies of exciton formation in the extended multi-orbital
Hubbard model in 3D.\@
A quantitative description of this binding phenomenon is essential for a consistent explanation of
optical and photoemission data and will provide important information on electronic correlations
in Hubbard systems.

\section*{Acknowledgement}

We thank I. Simons and M. Cwik for technical support, and T. Wagner,
D.I. Khomskii, G.S. Uhrig, E. Pavarini, O.K. Andersen, N. Bl\"{u}mer,
G. Khaliullin, T. M\"{o}ller, and S.V. Streltsov for fruitful
discussions. This work is supported by the DFG in SFB 608.

\section*{Appendix: Role of oxygen defects}
\label{defects}

\begin{figure}[tb]
  \includegraphics[width=0.9\columnwidth,clip]{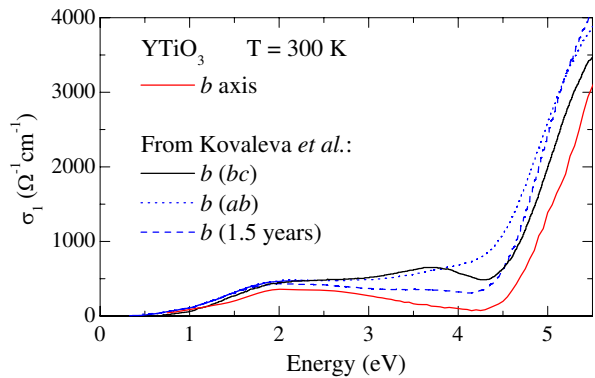}
  \caption{(Color online) Comparison of the optical conductivity for the $b$ axis of YTiO$_3$
  with spectra reported by Kovaleva {\it et al.},\cite{kovaleva07} which were determined from
  either the $ab$ or the $bc$ surface of a freshly polished sample, or from the $ab$ surface
  of a sample measured after 1.5 years.
   }
\label{fig:compKovaleva}
\end{figure}

Recently, Kovaleva {\it et al.}\cite{kovaleva07}  studied YTiO$_3$ by ellipsometry and reported
on complications which they attribute to oxygen defects arising from polar surfaces. They observed
peaks in $\sigma_1(\omega)$ at 1.95, 2.9, and 3.7\,eV.\@
The overall temperature dependence observed in Ref.\ \onlinecite{kovaleva07} is very weak,
showing a crossover at 100\,K, but no anomaly at $T_c$ within the experimental accuracy.
In the frequency range studied by us, the main effects of oxygen defects were identified as
(i) a shift of the fundamental absorption edge to lower frequencies,
(ii) an absorption band at about 0.8\,eV,
(iii) the absence of a pronounced minimum at 4.5\,eV, and
(iv) a shift of the onset of charge-transfer excitations to lower frequencies.
These shifts have been attributed to localized states at the edge of the
electronic bands. This sensitivity of the fundamental absorption edge to doping away from the half-filled
Mott insulator has been studied in Y$_{1-x}$Ca$_x$TiO$_3$ [\onlinecite{okimoto,taguchi}].
In Fig.\ \ref{fig:compKovaleva} we compare our data with the results of Ref.\ \onlinecite{kovaleva07}.
Our data show both the largest fundamental absorption edge and the largest onset frequency for
charge-transfer excitations, in combination with a pronounced minimum at 4.5\,eV.\@
We find good agreement with the spectrum of $\sigma_1(\omega)$ below the fundamental gap
determined in our group by transmittance and reflectance measurements on thin single
crystals\cite{rueckamp05a} (see Fig.\ \ref{fig:compareRueckamp}).
The transmittance clearly reveals bulk properties.
These data show no defect-induced absorption below the gap, the single weak feature observed at
about 0.3\,eV has been undoubtedly identified as a phonon-activated orbital
excitation.\cite{rueckamp05a,ulrich06,ulrich08}
Finally, we find clear anomalies in the vicinity of the magnetic ordering temperatures,
both in YTiO$_3$ and in SmTiO$_3$ (see Figs.\ \ref{fig:Y_Neff} and \ref{fig:SmNeff}).
The combination of all these observations provides strong evidence that we have observed the
intrinsic properties of YTiO$_3$.


\vspace*{-0.4cm}

\begin{thebibliography}{99}

\vspace*{-.4cm}

\expandafter\ifx\csname
natexlab\endcsname\relax\def\natexlab#1{#1}\fi
\expandafter\ifx\csname bibnamefont\endcsname\relax
  \def\bibnamefont#1{#1}\fi
\expandafter\ifx\csname bibfnamefont\endcsname\relax
  \def\bibfnamefont#1{#1}\fi
\expandafter\ifx\csname citenamefont\endcsname\relax
  \def\citenamefont#1{#1}\fi
\expandafter\ifx\csname url\endcsname\relax
  \def\url#1{\texttt{#1}}\fi
\expandafter\ifx\csname
urlprefix\endcsname\relax\def\urlprefix{URL }\fi
\providecommand{\bibinfo}[2]{#2}
\providecommand{\eprint}[2][]{\url{#2}}

\bibitem{kotliar04a}
G. Kotliar and D. Vollhardt, Physics Today \textbf{3}, 53 (2004).

\bibitem{imada98a}
M. Imada, A. Fujimori, and Y. Tokura, Rev.\ Mod.\ Phys.\
\textbf{70}, 1039 (1998).

\bibitem{tokura00}
Y. Tokura and N. Nagaosa, Science {\bf 288}, 462 (2000).

\bibitem{khaliullinrev}
G. Khaliullin,
Prog.\ Theor.\ Phys.\ Suppl.\ {\bf 160}, 155 (2005).
%% cond-mat/0510025.

\bibitem{khomskiirev}
D.I. Khomskii,
Physica Scripta {\bf 72}, CC8 (2005).
%% cond-mat/0508631.

\bibitem{vandenbrink95a}
J. van den Brink, M.B.J. Meinders, J. Lorenzana, R. Eder, and G. A. Sawatzky,
Phys.\ Rev.\ Lett.\ \textbf{75}, 4658 (1995).

\bibitem{gallagher97}
F.B. Gallagher and S. Mazumdar, Phys.\ Rev.\ B {\bf 56}, 15025 (1997).


\bibitem{neudert98}
R. Neudert, M. Knupfer, M.S. Golden, J. Fink, W. Stephan, K. Penc, N. Motoyama, H. Eisaki, and S. Uchida,
Phys.\ Rev.\ Lett.\ {\bf 81}, 657 (1998).

\bibitem{essler01}
F.H.L. Essler, F. Gebhard, and E. Jeckelmann, Phys.\ Rev.\ B {\bf 64}, 125119 (2001).

\bibitem{huebsch01}
A. H\"{u}bsch, J. Richter, C. Waidacher, K.W. Becker, and W. von der Linden,
Phys.\ Rev.\ B {\bf 63}, 205103 (2001).

\bibitem{jeckel03}
E. Jeckelmann, Phys.\ Rev.\ B {\bf 67}, 075106 (2003).

\bibitem{moskvin03}
A.S. Moskvin, J. M\'{a}lek, M. Knupfer, R. Neudert, J. Fink, R. Hayn, S.-L. Drechsler,
N. Motoyama, H. Eisaki, and S. Uchida, Phys.\ Rev.\ Letters {\bf 91}, 037001 (2003).

\bibitem{kim04}
Y.-J. Kim, J.P. Hill, H. Benthien, F.H.L. Essler, E. Jeckelmann, H.S. Choi, T.W. Noh,
N. Motoyama, K.M. Kojima, S. Uchida, D. Casa, and T. Gog,
Phys.\ Rev.\ Lett.\ {\bf 92}, 137402 (2004).

\bibitem{matsueda05}
H. Matsueda, T. Tohyama, and S. Maekawa, Phys.\ Rev.\ B {\bf 71}, 153106 (2005).

\bibitem{ono04}
M. Ono, K. Miura, A. Maeda, H. Matsuzaki, H. Kishida, Y. Taguchi, Y. Tokura, M. Yamashita, and
H. Okamoto, Phys.\ Rev.\ B {\bf 70}, 085101 (2004).

\bibitem{ono05}
M. Ono, H. Kishida, and H. Okamoto, Phys.\ Rev.\ Lett.\ {\bf 95}, 087401 (2005).

\bibitem{kishida}
H. Kishida, H. Matsuzaki, H. Okamoto, T. Manabe, M. Yamashita, Y. Taguchi, and Y. Tokura,
Nature {\bf 405}, 929 (2000).

\bibitem{clarke}
D.G. Clarke, Phys. Rev. B {\bf 48}, 7520 (1993).

\bibitem{wang}
Y.Y. Wang, F.C. Zhang, V.P. Dravid, K.K. Ng, M.V. Klein, S.E. Schnatterly, and
L.L. Miller, Phys. Rev. Lett. {\bf 77}, 1809 (1996).

\bibitem{zhangng}
F.C. Zhang and K.K. Ng, Phys. Rev. B {\bf 58}, 13520 (1998).

\bibitem{wrobel}
P. Wrobel and R. Eder, Phys.\ Rev.\ B {\bf 66}, 035111 (2002).

\bibitem{kuzian}
R.O. Kuzian, R. Hayn, and A.F. Barabanov, Phys.\ Rev.\ B {\bf 68}, 195106 (2003).

\bibitem{simon}
M.E. S\'{\i}mon, A.A. Aligia, C.D. Batista, E.R. Gagliano, and
F. Lema, Phys. Rev. B {\bf 54}, R3780 (1996).

\bibitem{hanamura}
E. Hanamura, N.T. Dan, and Y. Tanabe, Phys.\ Rev.\ B {\bf 62}, 7033 (2000);
J. Phys.: Condens. Matter {\bf 12}, L345 (2000).

\bibitem{moskvin}
A.S. Moskvin, R. Neudert, M. Knupfer, J. Fink, and R. Hayn,
Phys. Rev. B {\bf 65}, 180512(R) (2002).

\bibitem{gomi}
H. Gomi, A. Takahashi, T. Ueda, H. Itoh, and M. Aihara, Phys.\ Rev.\ B {\bf 71}, 045129 (2005).

\bibitem{itoh}
H. Itoh, A. Takahashi, and M. Aihara, Phys.\ Rev.\ B {\bf 73}, 075110 (2006).

\bibitem{collart}
E. Collart, A. Shukla, J.-P. Rueff, P. Leininger, H. Ishii, I. Jarrige,
Y.Q. Cai, S.-W. Cheong, and G. Dhalenne, Phys. Rev. Lett. {\bf 96}, 157004 (2006).

\bibitem{ellis}
D.S. Ellis, J.P. Hill, S. Wakimoto, R.J. Birgeneau, D. Casa, T. Gog, and Y.-J. Kim,
Phys.\ Rev.\ B {\bf 77}, 060501(R) (2008).


\bibitem{hirsch}
J.E. Hirsch Phys.\ Rev.\ Lett.\ \textbf{59}, 228 (1987);
Science \textbf{295}, 2226 (2002).

\bibitem{molegraaf02a}
H.J.A. Molegraaf, C. Presura, D. van der Marel, P.H. Kes, and M. Li,
Science \textbf{295}, 2239 (2002).

\bibitem{akimitsu01a}
J. Akimitsu, H. Ichikawa, N. Eguchi, T. Miyano, M. Nishi, and K. Kakurai,
J.\ Phys.\ Soc.\ Jpn.\ \textbf{70}, 3475 (2001).

\bibitem{iga04a}
F. Iga, M. Tsubota, M. Sawada, H.B. Huang, S. Kura, M. Takemura, K. Yaji, M. Nagira, A. Kimura, T. Jo,
T. Takabatake, H. Namatame, and M. Taniguchi,
Phys.\ Rev.\ Lett.\ \textbf{93}, 257207 (2004).

\bibitem{komarek}
A.C. Komarek, H. Roth, M. Cwik, W.-D. Stein, J. Baier, M. Kriener, F. Bour\'{e}e,
T. Lorenz, and M. Braden, Phys.\ Rev.\ B {\bf 75}, 224402 (2007).

\bibitem{kovaleva04a}
N.N. Kovaleva, A.V. Boris, C. Bernhard, A. Kulakov, A. Pimenov, A.M. Balbashov, G. Khaliullin,
and B. Keimer, Phys.\ Rev.\ Lett.\ \textbf{93}, 147204 (2004).

\bibitem{oles05a}
A.M. Oles, G. Khaliullin, P. Horsch, and L.F. Feiner,
Phys.\ Rev.\ B {\bf 72}, 214431 (2005).

\bibitem{lee05a}
J.~S. Lee, M.~W. Kim, and T.~W. Noh, New J.\ Phys. \textbf{7}, 147 (2005).

\bibitem{goessling08}
A. G\"{o}ssling, M.W. Haverkort, M. Benomar, Hua Wu, D. Senff, T. M\"{o}ller, M. Braden, J.A. Mydosh,
and M. Gr\"{u}ninger, Phys.\ Rev.\ B {\bf 77}, 035109 (2008).

\bibitem[{\citenamefont{Ahn and Millis}(2000)}]{ahn00}
\bibinfo{author}{\bibfnamefont{K.~H.} \bibnamefont{Ahn}} \bibnamefont{and}
  \bibinfo{author}{\bibfnamefont{A.~J.} \bibnamefont{Millis}},
  \bibinfo{journal}{Phys. Rev. B} \textbf{\bibinfo{volume}{61}},
  \bibinfo{pages}{13545} (\bibinfo{year}{2000}).

\bibitem[{\citenamefont{Tobe et~al.}(2001)\citenamefont{Tobe, Kimura, Okimoto,
  and Tokura}}]{tobe01a}
\bibinfo{author}{\bibfnamefont{K.}~\bibnamefont{Tobe}},
  \bibinfo{author}{\bibfnamefont{T.}~\bibnamefont{Kimura}},
  \bibinfo{author}{\bibfnamefont{Y.}~\bibnamefont{Okimoto}}, \bibnamefont{and}
  \bibinfo{author}{\bibfnamefont{Y.}~\bibnamefont{Tokura}},
  \bibinfo{journal}{Phys. Rev. B} \textbf{\bibinfo{volume}{64}},
  \bibinfo{pages}{184421} (\bibinfo{year}{2001}).

\bibitem[{\citenamefont{Lee et~al.}(2002)\citenamefont{Lee, Lee, Noh, Oh, Yu,
  Nakatsuji, Fukazawa, and Maeno}}]{lee02a}
\bibinfo{author}{\bibfnamefont{J.~S.} \bibnamefont{Lee}},
  \bibinfo{author}{\bibfnamefont{Y.~S.} \bibnamefont{Lee}},
  \bibinfo{author}{\bibfnamefont{T.~W.} \bibnamefont{Noh}},
  \bibinfo{author}{\bibfnamefont{S.-J.} \bibnamefont{Oh}},
  \bibinfo{author}{\bibfnamefont{J.}~\bibnamefont{Yu}},
  \bibinfo{author}{\bibfnamefont{S.}~\bibnamefont{Nakatsuji}},
  \bibinfo{author}{\bibfnamefont{H.}~\bibnamefont{Fukazawa}}, \bibnamefont{and}
  \bibinfo{author}{\bibfnamefont{Y.}~\bibnamefont{Maeno}},
  \bibinfo{journal}{Phys. Rev. Lett.} \textbf{\bibinfo{volume}{89}},
  \bibinfo{pages}{257402} (\bibinfo{year}{2002}).

\bibitem[{\citenamefont{Kim et~al.}(2004)\citenamefont{Kim, Lee, Noh, Yu, and
  Moritomo}}]{kim04a}
\bibinfo{author}{\bibfnamefont{M.~W.} \bibnamefont{Kim}},
  \bibinfo{author}{\bibfnamefont{Y.~S.} \bibnamefont{Lee}},
  \bibinfo{author}{\bibfnamefont{T.~W.} \bibnamefont{Noh}},
  \bibinfo{author}{\bibfnamefont{J.}~\bibnamefont{Yu}}, \bibnamefont{and}
  \bibinfo{author}{\bibfnamefont{Y.}~\bibnamefont{Moritomo}},
  \bibinfo{journal}{Phys. Rev. Lett.} \textbf{\bibinfo{volume}{92}},
  \bibinfo{pages}{027202} (\bibinfo{year}{2004}).

\bibitem[{\citenamefont{Rauer et~al.}(2006)\citenamefont{Rauer, R\"{u}bhausen, and
  D\"{o}rr}}]{rauer}
\bibinfo{author}{\bibfnamefont{R.}~\bibnamefont{Rauer}},
  \bibinfo{author}{\bibfnamefont{M.}~\bibnamefont{R\"{u}bhausen}},
  \bibnamefont{and} \bibinfo{author}{\bibfnamefont{K.}~\bibnamefont{D\"{o}rr}},
  \bibinfo{journal}{Phys. Rev. B} \textbf{\bibinfo{volume}{73}},
  \bibinfo{pages}{092402} (\bibinfo{year}{2006}).

\bibitem[{\citenamefont{Miyasaka et~al.}(2002)\citenamefont{Miyasaka, Okimoto,
  and Tokura}}]{miyasaka02}
\bibinfo{author}{\bibfnamefont{S.}~\bibnamefont{Miyasaka}},
  \bibinfo{author}{\bibfnamefont{Y.}~\bibnamefont{Okimoto}}, \bibnamefont{and}
  \bibinfo{author}{\bibfnamefont{Y.}~\bibnamefont{Tokura}},
  \bibinfo{journal}{J. Phys. Soc. Jpn.} \textbf{\bibinfo{volume}{71}},
  \bibinfo{pages}{2086} (\bibinfo{year}{2002}).

\bibitem[{\citenamefont{Tsvetkov et~al.}(2004)\citenamefont{Tsvetkov, Mena, van
  Loosdrecht, van~der Marel, Ren, Nugroho, Menovsky, Elfimov, and
  Sawatzky}}]{tsvetkov04a}
\bibinfo{author}{\bibfnamefont{A.~A.} \bibnamefont{Tsvetkov}},
  \bibinfo{author}{\bibfnamefont{F.~P.} \bibnamefont{Mena}},
  \bibinfo{author}{\bibfnamefont{P.~H.~M.} \bibnamefont{van Loosdrecht}},
  \bibinfo{author}{\bibfnamefont{D.}~\bibnamefont{van~der Marel}},
  \bibinfo{author}{\bibfnamefont{Y.}~\bibnamefont{Ren}},
  \bibinfo{author}{\bibfnamefont{A.~A.} \bibnamefont{Nugroho}},
  \bibinfo{author}{\bibfnamefont{A.~A.} \bibnamefont{Menovsky}},
  \bibinfo{author}{\bibfnamefont{I.~S.} \bibnamefont{Elfimov}},
  \bibnamefont{and} \bibinfo{author}{\bibfnamefont{G.~A.}
  \bibnamefont{Sawatzky}}, \bibinfo{journal}{Phys. Rev. B}
  \textbf{\bibinfo{volume}{69}}, \bibinfo{pages}{075110}
  (\bibinfo{year}{2004}).

\bibitem[{\citenamefont{Khaliullin et~al.}(2004)\citenamefont{Khaliullin,
  Horsch, and Ole\'{s}}}]{khaliullin04a}
\bibinfo{author}{\bibfnamefont{G.}~\bibnamefont{Khaliullin}},
  \bibinfo{author}{\bibfnamefont{P.}~\bibnamefont{Horsch}}, \bibnamefont{and}
  \bibinfo{author}{\bibfnamefont{A.~M.} \bibnamefont{Ole\'{s}}},
  \bibinfo{journal}{Phys. Rev. B} \textbf{\bibinfo{volume}{70}},
  \bibinfo{pages}{195103} (\bibinfo{year}{2004}).


\bibitem{RothPhD}
H. Roth, PhD thesis, University of Cologne (2008).

\bibitem{schubert96a}
M.~Schubert, Phys.\ Rev.\ B \textbf{53}, 4265 (1996).

\bibitem{goesslingthesis}
A. G\"{o}ssling, Ph.D. thesis, University of Cologne (2007);
{\mbox http://nbn-resolving.de/urn:nbn:de:hbz:38-21379}
%% {\mbox http://kups.ub.uni-koeln.de/volltexte/2007/2137 }

\bibitem{arima}
T. Arima, Y. Tokura, and J.B. Torrance,
Phys.\ Rev.\ B {\bf 48}, 17006 (1993).

\bibitem{rueckamp05a}
R. R\"{u}ckamp, E. Benckiser, M.W. Haverkort, H. Roth, T. Lorenz, A. Freimuth,
L. Jongen, A. M\"{o}ller, G. Meyer, P. Reutler, B. B\"{u}chner, A. Revcolevschi,
S.-W. Cheong, C. Sekar, G. Krabbes, and M. Gr\"{u}ninger,
New J.\ Phys.\ \textbf{7}, 144 (2005).

\bibitem{kovaleva07}
N.N. Kovaleva, A.V. Boris, P. Yordanov, A. Maljuk, E. Br\"{u}cher, J. Strempfer, M. Konuma, I. Zegkinoglou,
C. Bernhard, A.M. Stoneham, and B. Keimer,
Phys.\ Rev.\ B \textbf{76}, 155125 (2007).

\bibitem{fujimori92a}
A. Fujimori, I. Hase, H. Namatame, Y. Fujishima, Y. Tokura, H. Eisaki, S. Uchida,
K. Takegahara, and F.M.F. de Groot, Phys.\ Rev.\ Lett.\ \textbf{69}, 1796 (1992).

\bibitem{bocquet96a}
A.E. Bocquet, T. Mizokawa, K. Morikawa, A. Fujimori, S.R. Barman, K. Maiti,
D.D. Sarma, Y. Tokura, and M. Onoda,
Phys.\ Rev.\ B \textbf{53}, 1161 (1996).

\bibitem{morikawa96a}
K. Morikawa, T. Mizokawa, A. Fujimori, Y. Taguchi, and Y. Tokura,
 Phys.\ Rev.\ B \textbf{54}, 8446 (1996).

\bibitem{Uav}
For a comparison of different results one needs to distinguish
between $U\! = \! F^0 + \frac{4}{49} F^2 + \frac{36}{441} F^4$ (both
electrons occupy the {\em same} real orbital) and $U_{av}\! = \!
F^0-\frac{14}{441}(F^2+F^4)$ (averaged over all multiplets) with the
Slater integrals $F^0$, $F^2$, and $F^4$. The value of
$U_{av}\cong4$\,eV reported in Ref.\ \onlinecite{bocquet96a} corresponds
to $U \approx 5.3$\,eV.\@ However, this value is based on a TiO$_6$
configuration-interaction cluster model including covalency. Due to
screening effects by O$_{2p}$ orbitals, the effective value of $U$
is somewhat smaller, in good agreement with $U$=4.5\,eV chosen in
Fig.\ \ref{fig:levels}.


\bibitem{tanabesugano}
S. Sugano, Y. Tanabe, and H. Kamimura,
{\em Multiplets of Transition-Metal Ions in Crystals} (Academic Press, New York, 1970).

\bibitem{higuchi}
T. Higuchi, T. Tsukamoto, M. Watanabe, M.M. Grush, T.A. Callcott, R.C. Perera,
D.L. Ederer, Y. Tokura, Y. Harada, Y. Tezuka, and S. Shin,
Phys.\ Rev.\ B {\bf 60}, 7711 (1999).

\bibitem{ulrich08}
C. Ulrich, G. Ghiringhelli, A. Piazzalunga, L. Braicovich, N.B. Brookes, H. Roth, T. Lorenz,
and B. Keimer, Phys.\ Rev.\ B {\bf 77}, 113102 (2008).

\bibitem{schmitz05c}
R. Schmitz, O. Entin-Wohlman, A. Aharony, and E. M\"{u}ller-Hartmann,
Annalen der Physik \textbf{16}, 425 (2007);
Annalen der Physik \textbf{14}, 626 (2005).

\bibitem{ulrich06}
C. Ulrich, A. G\"{o}ssling, M. Gr\"{u}ninger, M. Guennou, H. Roth, M. Cwik, T. Lorenz, G. Khaliullin,
and B. Keimer, Phys.\ Rev.\ Lett.\ {\bf 97}, 157401 (2006).

\bibitem{t2gsplitting}
The $t_{2g}$ splitting $\Delta_{t2g}$ is relevant for a precise
determination of $U$, since the lowest non-excitonic excitation
(peak B) is expected at about $U-3J_H+\Delta_{t2g}$.

\bibitem{pavarini05a}
E. Pavarini, A. Yamasaki, J. Nuss, and O.K. Andersen,
New J.\ Phys.\ \textbf{7}, 188 (2005).

\bibitem{craco07}
L. Craco, S. Leoni, M.S. Laad, and H. Rosner, Phys.\ Rev.\ B {\bf 76}, 115128 (2007).

\bibitem{arita}
M. Arita, H. Sato, M. Higashi, K. Yoshikawa, K. Shimada, M. Nakatake, Y. Ueda,
H. Namatame, M. Taniguchi, M. Tsubota, F. Iga, and T. Takabatake,
Phys.\ Rev.\ B {\bf 75}, 205124 (2007).

\bibitem{magnon}
In Mott-Hubbard insulators, the particle and the hole may reside on the
{\em same} site, albeit with different spin or orbital quantum numbers.
Such a very strongly bound ``exciton'' corresponds to a
magnon or an orbiton, excitations within the spin or orbital
channel. There is no doubly occupied site and one does not have
to pay the energy $U$, in contrast to the excitations discussed here.

\bibitem{streltsov05a}
S.V. Streltsov, A.S. Mylnikova, A.O. Shorikov, Z.V. Pchelkina, D.I. Khomskii,
and V.I. Anisimov, Phys.\ Rev.\ B \textbf{71}, 245114 (2005).

\bibitem{solovyev06}
I.V. Solovyev, Phys.\ Rev.\ B {\bf 73}, 155117 (2006).

\bibitem{okimoto}
Y. Okimoto, T. Katsufuji, Y. Okada, T. Arima, and Y. Tokura,
Phys. Rev. B {\bf 51}, 9581 (1995).

\bibitem{taguchi}
Y. Taguchi, Y. Tokura, T. Arima, and F. Inaba, Phys. Rev. B {\bf 48}, 511 (1993).

\bibitem{mochizuki04a}
M. Mochizuki and M. Imada, New J. Phys. {\bf 6}, 154 (2004).

\bibitem{mochizuki04b}
M. Mochizuki and M. Imada, J. Phys. Soc. Jpn. {\bf 73}, 1833 (2004).

\bibitem{kubota}
M. Kubota, H. Nakao, Y. Murakami, Y. Taguchi, M. Iwama, and Y. Tokura,
Phys. Rev. B {\bf 70}, 245125 (2004).

\bibitem{harrison}
W.A. Harrison, {\it Elementary electronic structure} (World Scientific, 1999).


\end{thebibliography}
\end{document}